# NEW APPROACHES WITH CHORD IN EFFICIENT P2P GRID RESOURCE DISCOVERY

Jeyabharathi[1] and Pethalakshmi[2]

[1]Ph.D Scholar, Manonmaniam Sundaranar University, Tirunelveli
bharathi_guhan@yahoo.com
[2]Associate Professor and Head of Computer Science, M.V.M Government Arts College(W), Dindigul
pethalakshmi@yahoo.com

## ABSTRACT

Grid computing is a type of distributed computing which allows sharing of computer resources through Internet. It not only allows us to share files but also most of the software and hardware resources. An efficient resource discovery mechanism is the fundamental requirements for grid computing systems, as it supports resource management and scheduling of applications. Among various discovery mechanisms, Peer-to-Peer (P2P) technology witnessed rapid development and the key component for this success is efficient lookup applications of P2P. Chord is a P2P structural model widely used as a routing protocol to find resources in grid environment. Plenty of ideas are implemented by researchers to improve the lookup performance of chord protocol in Grid environment. In this paper, we discuss the recent researches made on Chord Structured P2P protocol and present our proposed methods in which we use the address of Recently Visited Node (RVN) and fuzzy technique to easily locate the grid resources by reducing message complexity and time complexity.

## KEYWORDS

Grid Resource Discovery, Recently-Visited-Node, Peer-to-Peer, Fuzzy Classification, Intrusion Detection

## 1. INTRODUCTION

"Grid is the future Internet", is the prime mantra of the present researchers and so whole world follows Grid. Grid applications are usually scientific with large number of users and dynamic resources. Because of the dynamic nature of Grid systems, it allows participants to join or leave the system at any time. To discover large number of dynamic resources an efficient, scalable and accurate discovery mechanisms are needed. Grid computing and P2P computing models share more features in common and P2P techniques and protocols can be used to implement scalable services and applications. The main reason for usingP2P techniques in Grid is that it supports scalability which is the key requirement of Grid systems. The two key services [1] of Grid managed by P2P techniques are membership management and resource discovery. The objective of a membership management services is adding a new node to the network and assigning this node a set of neighbour nodes. The resource discovery service is invoked by a node when it is needed to discover and use different types of resources.

Resource discovery in Grid is a process of locating proper resource candidates which are suitable for executing jobs within a reasonable time. Efficient usage of the right resources is the key component of success of the Grid systems. The characteristics of the Grid systems make the resource discovery a time consuming process which can decrease the performance of the whole system. Various methods have been proposed to solve the resource discovery problem in Grid systems. They are classified into three main categories [2].

Grid resource discovery process uses different classes of systems like centralized and hierarchical systems and agent based systems. Even though these methods have the advantage of Open Grid Service Architecture (OGSA), they suffer from scalability, reliability and false-positive problems respectively. On the other hand, agent based systems [3] are attractive in Grid systems because of their autonomy properties. They have capabilities to determine new migration sites according to their migration policies for the distribution of resource discovery queries, so that researchers adopted Peer-to-Peer (P2P) technology in Grid environment to solve these problems.

In recent years P2P systems have been the hottest research topic in a large distributed system. Since P2P based network approach may overcome the limitations of hierarchical and centralized methods, P2P techniques are especially used in resource discovery process. The self-organization, scalability and dynamicity are the inspiring features of P2P systems. As P2P network is a kind of distributed network, the nodes of P2P network share their own part of the hardware resources like processing power, storage capacity, network connectivity, RAM, virtual memory etc.,

P2P systems are mainly divided into Structured P2P networks, Unstructured P2P networks and Super-Peer systems [4]. Unstructured P2P resource discovery approaches handle the dynamicity of resources. The routing mechanism of unstructured approaches presents the Grid to scale. In super-peer based methods, flooding [5] mechanism is used which leads to single point of failures.

The structured P2P methods [6, 7] ensure the scalability of the system by involving all the nodes in the query processing. This ensures that all nodes in grid will have the equal load. The structured P2P networks based on DHT, which uses structured hash algorithm for hashing resources and node ids in the same space. The DHT based most well-known protocols are Chord [8], Pastry [9], Tapstry [10] and CAN [11]. Among these Chord algorithm is simple and easy to design and implement. Due to its simplicity, scalability and high efficiency Chord lookup protocol has been widely researched and applied in Grid environment especially in resource discovery.

In this paper, we present our new method for Resource Discovery and we compare this with our proposed method1 which we have already experimented. In our proposed method2, we restructure Chord in a different manner that nodes with more number of resources and unique resources are selected and considered for lookup process. Then a brief comparison is made between the two works and given clearly through Graphs and tables. This method reduces the time to search unnecessary nodes and end within minimum hops. It also reduces the required number of messages to locate the resources. Finally we conclude that application of feature selection technique plays a significant role in resource discovery process.

## 2. RELATED WORKS

### 2.1. Chord

Chord, CAN, Tapstry and Pastry are the original distributed hash table protocols. Chord protocol is proposed by Ion Stocia, Robert Morris, David Karger, Frans Kaashoek and Hari Balakrishnan, and was developed at MIT. Chord assigns each node a unique ID by hashing the node's IP address to a binary string of fixed length of m. It also hashes each key to an m-bit binary string. The circle can have IDs/Keys ranging from 0 to $2^m - 1$. The consistent hash function assigns each node and key an m-bit identifier using secured hash algorithm (SHA-1). In Chord, identifiers are ordered in an identifier circle modulo $2^m$. Each node has a successor and a predecessor. The successor to a node (or Key) is denoted by successor (K) and is the next node in the identifier circle in clock-wise direction. The predecessor is counter-clockwise.

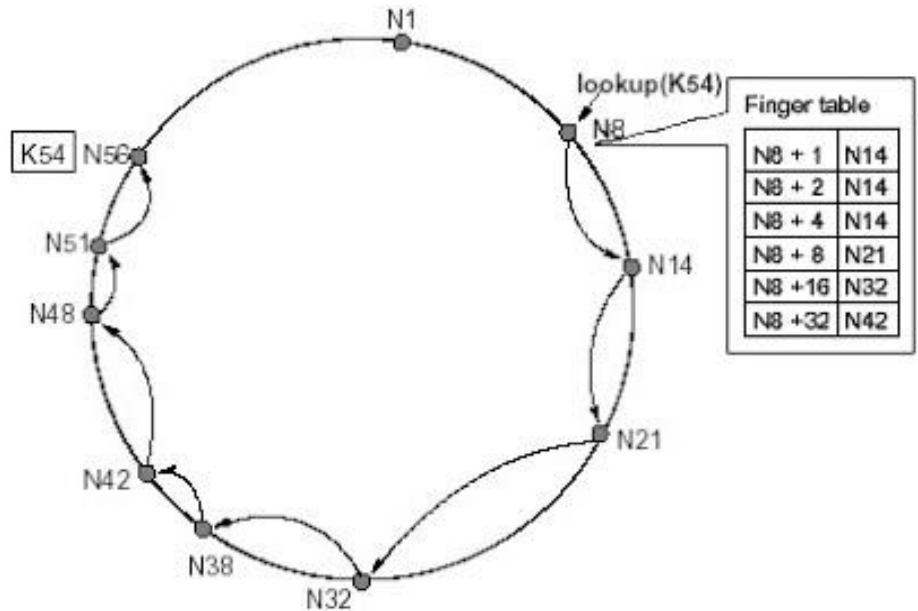

Figure 1. Chord Ring with 64 nodes

Figure 1 shows an identifier circle with m=6. The circle has ten active nodes and the identifier ranges from 0 to 63. In this example, the successor of identifier 1 is node 1, so the key 1 would be located at node 1. In the same way, key 29 is located at node 32, and key 44 at node 48. Consistent hashing is designed to let nodes enter and leave the network with minimal disruption. To maintain the consistent hashing mapping, when a node n joins the network, certain keys previously assigned to n's successor now are assigned to n. When node n leaves the network, all of its assigned keys are reassigned to n's successor. In the example given above, if a node were to join with identifier 37, it would capture the key with identifier 36 from the node with identifier 38.

Consistent Hashing allows nodes to enter and leave network freely with minimal disruption. To maintain consistent hashing, stabilization procedure is called periodically to change successor and predecessor of the node to join or leave. Every node maintains a finger table which contains addresses of m successor nodes. The $i^{th}$ entry of node is the address of successor of $((n + 2^{i-1}) \bmod 2^m)$. With the help of the finger table entries, each node which demands resource can easily locate the successor of the specific resource. The base Chord ring uses the following table to construct finger table.

Table 1. Finger Table of Chord

| Finger[k].start | $(n+2^{k-1}) \bmod 2^m$ |
|---|---|
| Finger[k].interval | [Finger[k].start, Finger [k+1].start) |
| Finger[k].node | successor (Finger[k].start) |

## 2.2. Role of Chord in Grid Resource Discovery-Existing Methods

Till recently, researchers have proposed many improved schemes to enhance routing efficiency of Chord Protocol. To deal with the routing delay and ignorance of physical topology information in P2P, [12] presents a topology-aware structured P2P system, named TB-Chord, which applies a suit of mechanisms to extend Chord to optimize the utilization of physical network topology and overlay network. The TB-Chord makes effective use of network topology structure during network routing. The result of experiments show that the TB-Chord, compared with traditional Chord, has obvious improvements in the routing delay and the hops of overlay networks. At the same time, TB-Chord does not need super node or plus layer which will pay more maintenance costs. Fan chao et al., [13] proposed BNN-Chord algorithm based on neighbors' neighbor (NN) chord. NN-Chord algorithm extends finger table using neighbors' neighbour link which is based on learn table. This table maintains information of the successor's successor node, which can reasonably increase finger density of routing table to find the neighbor node, which is more close to the object. The BNN-Chord algorithm effectively reduces the search path length and can improve the system performance.

Chunling Cheng et al [14] proposed advanced chord routing algorithm based on redundant information which replaced finger table with objective resource table. In base Chord the finger table contains redundant information which takes valuable and also increases search delay. In this paper, an objective resource table is established to solve the repeated search problem. Each node can establish links with many nodes and the search performance is greatly improved. To handle multi-attribute Multi-keyword fuzzy-matching queries with High recall ratio and load balancing, ZHAO Xiu-Mei et al., [15] proposed a new resource indexing model which is expanded from Chord and called MF-Chord. Yufeng Wang et al., [16] analyzed the of Chord algorithm to reconstruct the finger table in Chord, in which counter clockwise finger table is added to achieve resource queries in both directions, and the density of neighboring fingers is increased. Additionally, AB-Chord implements a new operation to remove the redundant fingers introduced by adding fingers in AB-Chord. In comparison with original Chord, new fingers are inserted into the middle of two neighboring fingers in clockwise and counter clockwise direction. AB-Chord improves query efficiency in terms of average lookup hops and average lookup delay. AB-Chord is further implemented as AB-Chord+ which extends periodic time of updating finger table and makes the joining and leaving nodes actively send updating messages. This method reduced the network bandwidth consumption. Huayun Yan et al., [17] modified the finger table of original Chord, and modified all the places which relate to the entry of the finger table, such as the node join, routing procedure etc. Huayan Yan et al., get a very perfect routing result through experiment and found that the routing hops is close to a constant number. The cost is more updating when joining a node and deleting a node in the system and nearly double the size of finger table. This system will be more effective in a comparatively steady condition. Miano Yuting et al., [18] presented Hot – Chord in which an inner chord is constructed dynamically by using the nodes, where the hot resources are located. Each inner node maintains an additional routing table, which records O (log $H_m$) other inner node's information and the mapping relationship between node's new Hash value and original Hash value. This algorithm decrease the inquiring space and average query hops and increase the query efficiency of the hot resources.

## 3. PROPOSED METHODS

### 3.1. Introduction

From section 2.2, it is clear that Chord protocol and modifications in finger table significantly improves the lookup performance of Resource discovery process. Since chord protocol searches only a single keyword at a time, SHA-1 function is highly suitable for hashing IP address of nodes and keys maintained by each node. We have proposed two methods in which we tried to improve the lookup performance of chord protocol. In our first method we use the address of Recently-Visited-Node id to search the resource which will find the key shortly if the previous path is useful for current search. In our second method we split the given chord ring into three sub-rings by applying fuzzy classification. The splitting is based on the number of resources each node maintains. Searching of resource is done simultaneously in all the three rings and resource is located in minimum hops. The results of the two methods are compared in terms of Number of messages, Number of hops and average communication time.

### 3.2. Proposed Method1

In this section, we present the modified chord protocol by adding an additional entry into the Base Chord finger table to store the location of Recent Visited Node (RVN). The next lookup will use this id to locate its key, if the match is found. The primary aim of any protocol is the efficient and fast lookup of nodes containing keys. Generally the performance of Chord like algorithm can be analyzed in terms of three metrics: the size of the finger table of every node, the number of hops a request needs to travel in the worst case, and the average number of hops. In the original Chord, when a node requests a key, it has to search its own finger table first and it may find the successor of the key in that table if the match is found. Otherwise, the node sends messages to other whose node id is less than or equal to the searched key and it may need a few hops to locate it. After locating the successor of the key, the result is returned to the node which started the search and the lookup process is successfully done. Our protocol modifies the finger table of the original Chord to add a new entry in the finger table which stores the recent visited node's id.

In RVN-Chord, we are using SHA1 hash function to map key and node id as in the original Chord. Our modified finger table stores the Recent Visited Node id which reserved the key of the previous lookup. This id will be stored in all the nodes of the Chord ring. This is identified by the variable RVN-id which is the first column entry of any finger table. This RVN-id is similar to the use of Recent Document list in our computer file system. By using the recent profile, the next search will be easy and we find the document's location quickly if the search matches with the recent document. Similarly in our proposed model, for every new search the node starts lookup by checking RVN-id of its finger table which was updated by the previous lookup. If the RVN-id matches the query, the process will succeed and it can directly locate the destination. If it is not exactly matched, the node next checks if the key is greater than the RVN-id. If it is true than the recent route id is in its destination path. Then it can jump to that id and continue its search from that location. Otherwise the normal Chord algorithm is invoked to find the successor of key. This type of lookup will reduce the number of hops, messages and communication time. Since Memory consumed by the finger table to store the RVN-id is very less in terms of bytes it will not be a big issue.

#### 3.2.1. RVN-Chord Algorithm

The following are the steps followed in our proposed method.
  i. Construction of finger table

   a. Add a new entry as follows.
         RVN.id = last visited node.id

   b. Finger[k].start         : $(n+2^{k-1}) \mod 2^m$
      Finger[k].interval      : [Finger[k].start, Finger[k+1].start)
      Finger[k].node          : successor (Finger[k].start)

  ii. Lookup of key using Recent_Visited_Node.id

   a. Check if key= Recent_Visited_Node.id  then
         Successor (key) = Recent_Visited_Node.id

Else

Check if key > Recent_Visited_Node.id then Jump to Recent_Visited_Node.id then continue the search.

iii. If step 2 fails, use the normal search as in base chord method.

iv. Call step 1(a) to update the finger table for every successful search.

v. When the visited node leaves the ring or any failure of that node occurs, set Recent_Visited_Node = successor(Recent_Visited_Node) and update all finger tables.

Figure 2 depicts a chord ring whose m is equal to 7(m is the bit size of the keywords or identifiers). The identifiers ranges from 0 to 127($2^7$-1).Here, we are comparing RVN Chord with the base Chord algorithm. The number of hops and messages are compared with both models for searching a few keys. Consider node N1 searches key 86 in the Chord ring. In normal Chord, in the process of node N1 searching for key 76, it first searches node N1, node N4 and then node N77. It needs 2 hops and 12 messages to find the object node reserving the key N76. The next search can be started by any node and let us consider the node N10 looking for key 90. The process starts from node N10 and moves to N45, N77 and finally N92. It needs 3 hops and 16 messages.

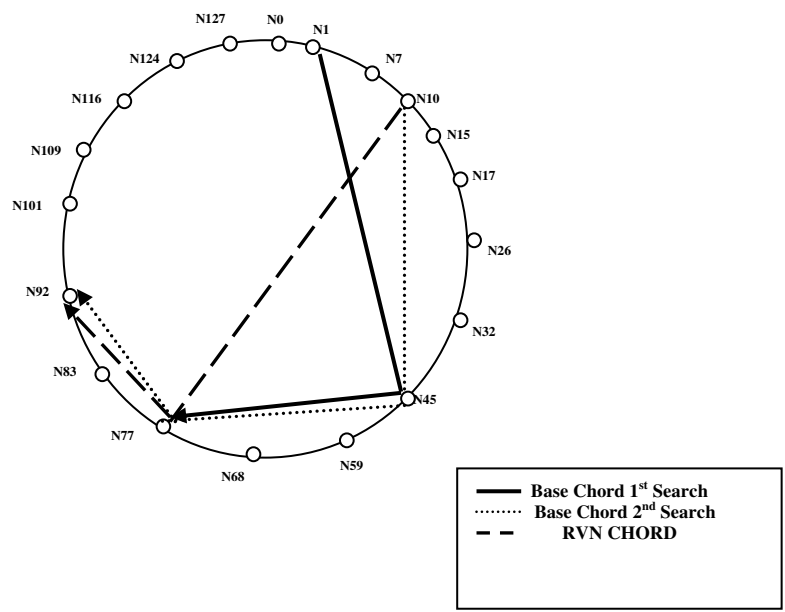

Figure 2. Chord Ring with 128 nodes

Now we apply RVN-Chord to find the successor of the key. Assume that the node N10 looks for the key 92. When the process is started RVN-Chord verifies its Recent Visited Node id for quick search, because it stored the last visited node in its finger table's first column. As per our previous example, the RVN-id is 77 (as in base chord). Node N10 first checks RVN.id for equality (if 90 = 77) and the match is not found. It then checks condition (N90 > N77) and it is true. So it jumps to N77. Next it searches the finger table of N77 and finds that N92 is the 4[th] finger of N77 which reserves the key 90. The lookup starts at node N10, moves to node N77 and locates the key at node N92. This process requires only 2 hops and 5 messages to locate the key whereas base Chord locates the key in 3 hops and 16 messages. The RVN-Chord algorithm performs well for the lookup of keys whose successor is equal to RVN.id or greater than RVN.id.

## 3.3. Proposed Method2

In this method we apply Fuzzy classification which classifies elements into a fuzzy set and its membership function is defined by the truth value of a fuzzy propositional function. The Chord is constructed initially with $2^m$ nodes. The ring is divided into three rings according to the basic Fuzzy- rule. Nodes with more than 66% of resources are grouped in HOTTEST RING. Nodes with resources between 34 to 65% are grouped into HOTTER RING and nodes with less than 34% resources are allocated to the HOT RING. The goal of this method is to create a model which includes all the necessary conditions to clearly differentiate each ring. Whenever a new node enters, the algorithm correctly predicts the corresponding ring and the node will join into the desired place. In this place Divide-and-Conquer learning concept is applied to split the nodes into subsets and this process is recursively executed for the subset of nodes.

Fuzzy numbers are fuzzy subsets of the real line. They have a peak or plateau with membership grade 1, over which the members of the universe are completely in the set. The membership function is increasing towards the peak and decreasing away from it. The following figure depicts the fuzzy membership function for three groups of nodes.

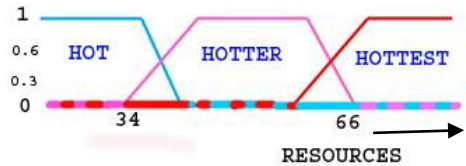

Figure 3: Fuzzy Membership Function for Three Rings

Now the $2^m$ nodes are grouped into three rings according to the classification procedure of decision tree algorithm with the help of basic fuzzy logic properties. The splitting is based on the number of resources each node has. In each ring, Ring-Head is selected based on the strength of the resources. Ring-Head is the entry point to each ring. If a query is started by the node of any ring, the Ring-Head checks its own ring and simultaneously sends the query to the Ring-Head of the remaining two Ring-Head. So the resources are simultaneously searched in all the three rings. The lookup process may successfully end in the ring which originates the query or any one of the two rings may find the resource. If the resource is located in the originated ring, the process is similar to the normal lookup process. Otherwise, the node which stores the key will return the result to its Ring-Head and in-turn the Ring-Head submits the result to the corresponding node via the head of the corresponding ring. As far as the lookup process is concerned, this method will find the key through parallel search and reduces the communication time dramatically. We named this method as FZ-Chord.

After splitting the rings, Resource Table is constructed by getting details of nodes from the finger table of each node. The entries of the table are node.id, number of resources and node status. Node status stores the status of resources of each node.id.

Table 2. The Structure of Resource Table

| Node.id | No. of Resources | Node Status |
|---------|------------------|-------------|

Entropy is used at the time of creating Resource Table to avoid duplicate entries of resources. If more than one node maintains the same set of resources, it will be handled by entropy method to put single entry about those resources. For example, if nodes N2, N4, N9 have RAM with 1GHZ speed, Resource Table will have a single entry for these three nodes. Entropy finds this redundancy and makes the Resource Table meaningful.

### 3.3.1. FZ-Chord Algorithm

The following are the steps followed by our proposed method.
  i. Construct a Chord ring with $2^m$ nodes.

  ii. Split the ring into three by applying Fuzzy classification. For splitting, we use the following steps

    a) If number of resources > 66% then include the corresponding nodes in the HOTTEST RING. Find the unique node, using Intruder detection system and include the node in HOTTEST RING.

    b) If number of resources are between 34% and 66% then include the corresponding nodes in the HOTTER RING.

    c) If number of resources < 34% then, include the nodes in HOT RING.

  iii. Ring-Head is assigned to each ring and it is the strongest node of the ring which stores more resources.

  iv. Construct Resource Table by getting node details from finger table.

  v. Resource Table is created once and parallel instances are distributed to all the nodes in all the rings.

### 3.3.2. EXAMPLE ANALYSIS

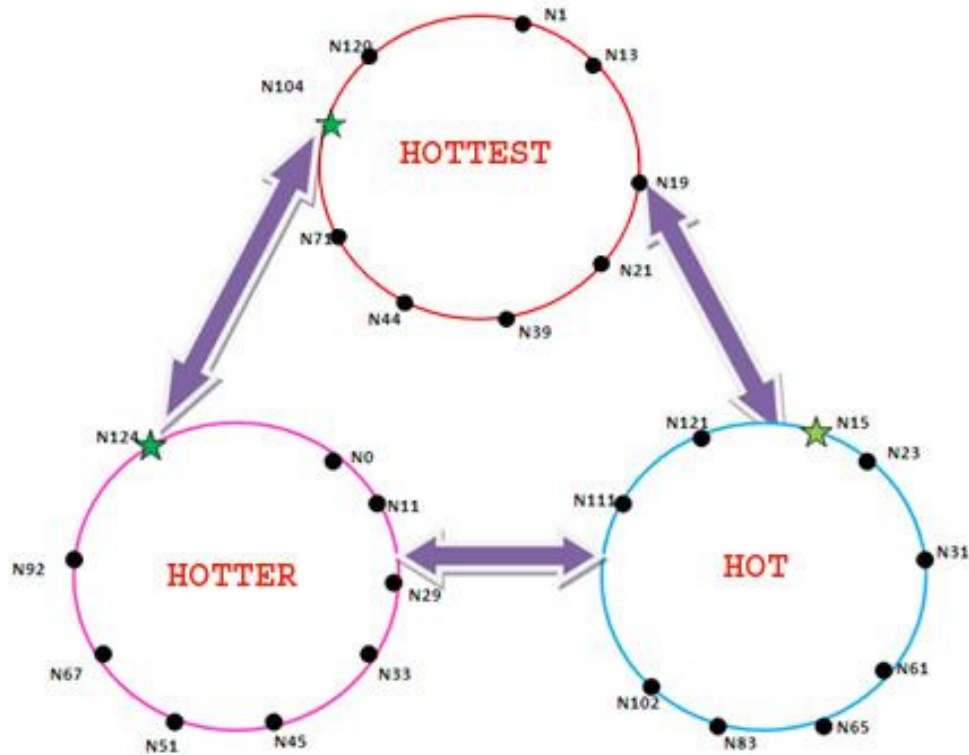

Figure 4. Three Chord Rings after Fuzzy classification

Chord Ring with $2^7$ (128) nodes is divided into three Rings as shown in figure 4.

Example 1:

Node 44 searches the key 83 which is in the HOT RING, the lookup process follows the following steps.

**Step 1:** Node 44 searches its own Resource Table and simultaneously it sends the request to its Ring-Head.

**Step 2:** Now Ring-Head of HOTTEST RING sends request to the Ring-Heads of Hotter Ring and Hot Ring.

**Step 3:** The search process is simultaneously going on with the help of Ring-Heads.

**Step 4:** Key 83 is located in the HOT RING and the corresponding Ring-Head sends the result to Node 44 through the Ring-Head of HOTTEST RING.

Example 2:

Node 67 searches key 10 which is actually located in HOTTEST RING. The search process is as follows

**Step 1:** Node 67 searches its own Resource Table and simultaneously it sends the request to its Ring-Head.

**Step 2:** Now Ring-Head of HOTTER RING sends a query request to the Ring-Heads of HOTTEST RING and HOT RING.

**Step 3:** The search process is simultaneously going on with the help of Ring-Heads.

**Step 4:** Key 10 is located in the HOTTEST RING and the corresponding Ring-Head sends the result to Node 67 through the Ring-Head of HOTTER RING.

### 3.3.3. Resource Based Intruder Detection (RID)

Node with unique resources is also included in the HOTTEST RING. For this purpose intrusion-detection method is used. Intrusion Detection is the act of detecting actions that attempt to compromise the confidentiality, integrity or availability of a resource. More specifically, the goal of intrusion detection is to identify entities attempting to subvert in-place security controls. The Common types of Intrusion Detection are Network Based (Network IDS), Host Based (HIDS) and Resource Based Intruder Detection (RID).

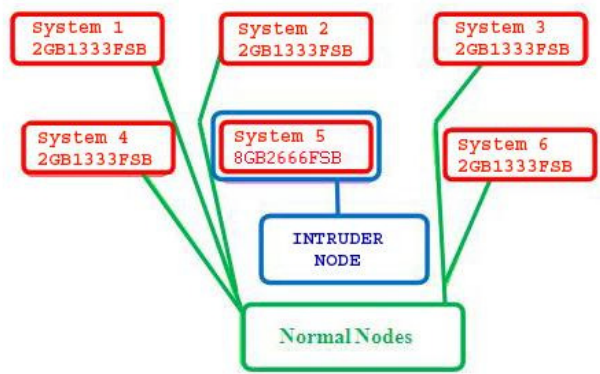

Figure 5. An Example of Intrusion Detection

In a typical network, many of the nodes share the same system specifications. If there is some node with odd configuration, then that is probably an intruder. Any entry level intruder detection system will identify this and raise an alarm. If the node is authorized, then the system will be instructed to treat the intruder detection system alert as FALSE Alarm. But in many cases, the node will be discarded by the network to ensure security. So Any Elementary Intruder detection System can be used to filter nodes with odd resource.

Figure 5 explains the typical network with Node characteristics and the Application of RID. System 5 is identified as unique node (Intruder node) which stores the unique resource. The RID algorithms classify nodes with unique resources or significant differences in resource. This can be used for easy classification of resources. Unique nodes are nodes with special resources which are not available in any of the nodes in the ring. These nodes are also called as intruder nodes. Intruder nodes are identified and incorporated in the HOTTEST RING. The Ring-Head maintains the location of unique node in its Resource Table.

## 4. SIMULATION EXPERIMENTS AND ANALYSIS

In section 3, we have a detailed discussion on our two proposed methods. In the first method previous history of lookup process is followed to find the resource and it will be quiet fast if the previous path is used by most of the lookup processes. In our second method, we use Fuzzy classification to split the Chord into multiple Rings and the search process is done simultaneously in all the Rings.

Table 3. Average number of hops, messages and communication time in Chord, RVN-Chord and FZ-Chord

| Messages : | Nodes | | | | | | | |
|---|---|---|---|---|---|---|---|---|
| | **256** | **512** | **1024** | **2048** | **4096** | **8192** | **16384** | **32768** |
| **Chord** | 20.4 | 42.1 | 64.8 | 87.9 | 136.6 | 149.7 | 179.3 | 219.8 |
| **RVN Chord** | 10.7 | 27.8 | 28.7 | 33.9 | 37.9 | 43.6 | 67.0 | 75.5 |
| **FZ Chord** | 5.0 | 19.5 | 19.4 | 21.9 | 27.8 | 31.0 | 53.0 | 66.3 |
| **Hops:** | | | | | | | | |
| **Chord** | 3.9 | 6.8 | 7.4 | 6.4 | 8.6 | 8.3 | 9.4 | 9.3 |
| **RVN Chord** | 3.8 | 4.4 | 5.0 | 4.5 | 4.6 | 5.8 | 6.3 | 7.5 |
| **FZ Chord** | 3.0 | 4.0 | 4.6 | 4.0 | 4.0 | 5.0 | 6.0 | 7.0 |
| **Communication Time:** | | | | | | | | |
| **Chord** | 1026.5 | 1124.7 | 1212.7 | 1265.7 | 1416.7 | 1521.8 | 2028.4 | 2846.8 |
| **RVN Chord** | 770.4 | 835.8 | 962.6 | 1065.1 | 1141.9 | 1223.4 | 1549.7 | 2036.3 |
| **FZ Chord** | 461.4 | 573.3 | 690.3 | 798.8 | 926.2 | 1065.0 | 1144.7 | 1267..4 |
| **Memory Consumed:** | | | | | | | | |
| **Chord** | 933.3 | 1069.7 | 3277.1 | 6417.4 | 12513 | 23964.5 | 55884.9 | 80739.6 |
| **RVN Chord** | 954.6 | 1090.1 | 3359.9 | 6551.6 | 12634.6 | 24166.4 | 56344.4 | 82384.9 |
| **FZ Chord** | 2174.5 | 4416.4 | 17965.4 | 29321.3 | 36811.7 | 60196.9 | 90615.3 | 100673.6 |

The two methods proposed in this paper are run in the NSC_SE simulator with N=256, 516, 1024, 2048, 4096, 8192, 16384 and 32768. The comparisons among base Chord, proposed method1 (RVN-Chord) and proposed method2 (Fuzzy classification) are made in terms of average number of hops, average number of messages, average communication time and memory consumed. As shown in table 2, average number of hops and average communication time are highly reduced in proposed method 2, which uses fuzzy classification. Hence we conclude that proposed method2 performs better than base Chord and proposed method1.

Figure 6, 7, 8 and 9 shows the average communication time, number of hops and message per peer for Chord, RVN-Chord and FZ-Chord with changes of total number of nodes in P2P systems.

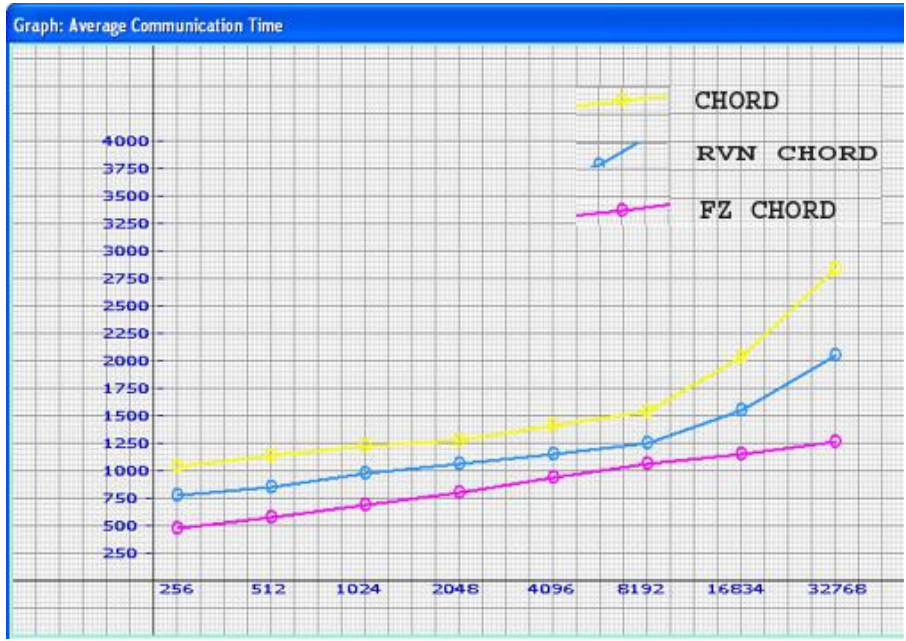

Figure 6. Average Communication Time

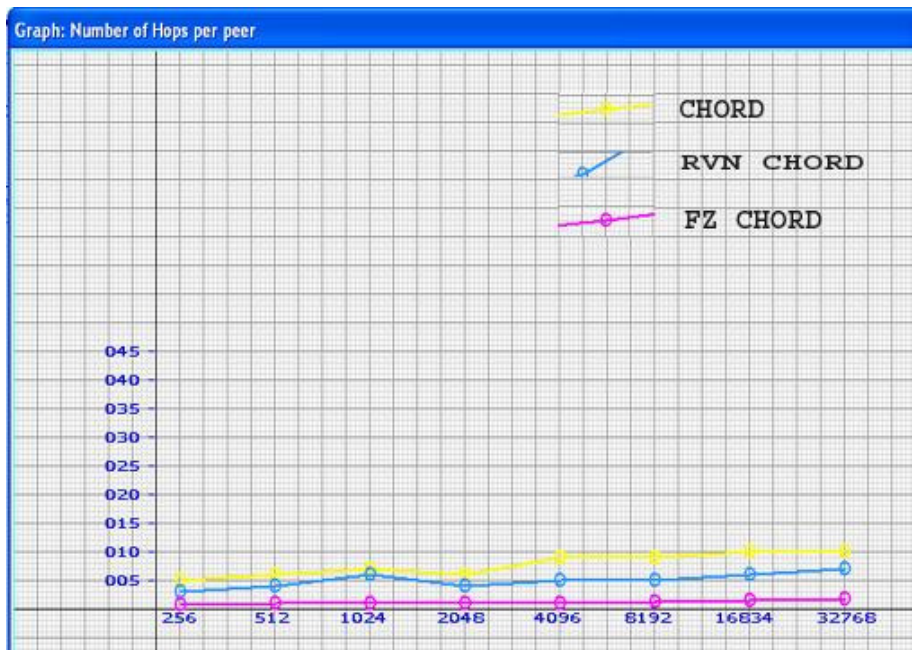

Figure 7. Number of Hops per Peer

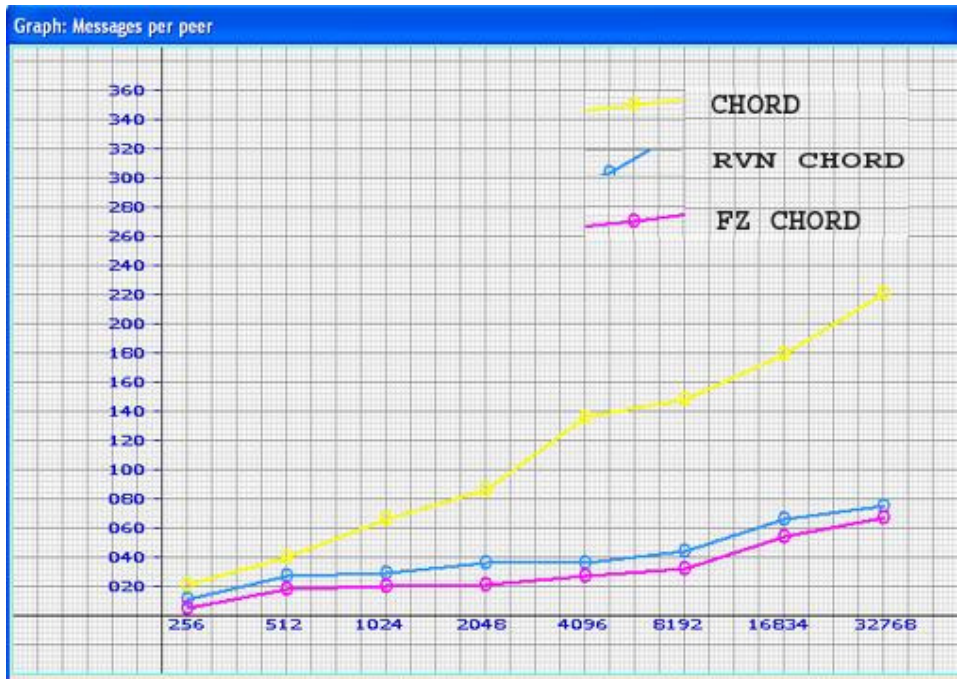

Figure 8. Message per Peer

As per the simulation result, RVN-Chord and FZ-Chord consumes more memory while using increased number of nodes. Proposed method1 takes little more memory to store the RVN.id which is a single column entry in the Finger table. But in fuzzy classification, Resource Table is maintained by all the nodes of all the Rings. So this method takes more memory than base Chord and RVN-Chord. Since memory is measured in terms of bytes, it will be not considered as a big issue.

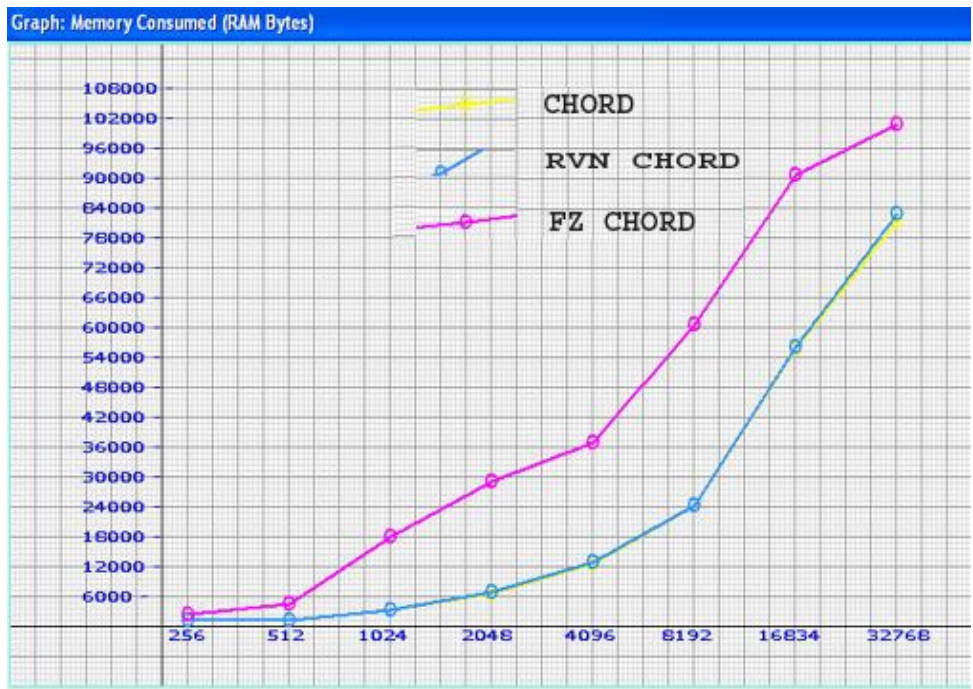

Figure 9. Memory Consumed

# 5. CONCLUSION

As more and more resources appear in grids, there is an increasing need to discover these resources effectively and efficiently. Our common aim is to efficiently locate the resources in Grid environment by using P2P techniques. Since Chord is the appropriate choice for single keyword search, we have experimented chord with two different systems. In the first method, previous history of lookup process is used to find the resources quickly if the previous path is used by most of the lookup processes. In our second method, we use Fuzzy classification to split the Chord into multiple Rings and the search process is done simultaneously in all the rings. From the simulation results, we conclude that the two proposed methods effectively reduce the required number of hops, messages and communication time. But the methods need a little more memory to store the previous history and Resource table respectively than the base Chord. It will not be an issue as memory is measured in terms of RAM-Bytes.

## Author Profile

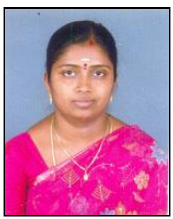 **C.Jeyabharathi,** received the Bachelor of Computer Science (B.Sc.,) degree from Madurai Kamaraj University, Madurai, TN, India, in 1997 and the Master of Computer Applications (M.C.A.) from the same University, in 2001. She has received the Master of Philosophy in Computer Science from Mother Teresa Women's University, Kodaikanal, Tamilnadu, India, in 2006. She is currently pursuing her Ph.D. degree with the Department of Computer Science, Manonmaniam Sundaranar University, Tirunelveli,TN, India. Her research interests include grid computing and data mining.

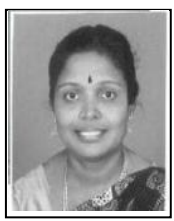 **Pethalakshmi Annamalai**, received the Master of Computer Science from Alagappa University, Karaikudi, TN, India, in 1988 and received the Master of Philosophy in Computer Science from Mother Teresa Women's University, Kodaikanal, TN, India, in 2000. She has received her Ph.D. Degree from Department of Computer Science, Mother Teresa Women's University, Kodaikanal, TN, India, in 2008. Currently she is working as Associate Professor and Head, Department of Computer Science, M.V.M. Govt. Arts College (w), Dindigul, TN, India. Her areas of interests include fuzzy, rough set, neural network and grid computing.